\documentclass[fleqn,usenatbib]{mnras}

\usepackage{newtxtext,newtxmath}

\usepackage[T1]{fontenc}

\DeclareRobustCommand{\VAN}[3]{#2}
\let\VANthebibliography\thebibliography
\def\thebibliography{\DeclareRobustCommand{\VAN}[3]{##3}\VANthebibliography}

\usepackage{graphicx}	

\title[New technique to select fast-quenching galaxies]{New technique to select recent fast-quenching galaxies at $\bf z\sim2$ using the optical colors}

\author[M. Kubo et al.]{
Mariko Kubo$^{1,2}$\thanks{E-mail: m.kubo@astr.tohoku.ac.jp},
Tohru Nagao$^{2}$,
Hisakazu Uchiyama$^{2,3}$,
Takuji Yamashita$^{3}$,
Yoshiki Toba$^{2,3,4}$,
\newauthor
Masaru Kajisawa$^{2}$,
Yuta Yamamoto$^{2}$
\\
$^{1}$Astronomical Institute, Tohoku University, 6-3, Aramaki Aoba, Aoba-ku, Sendai, Miyagi, Japan, 980-8578\\
$^{2}$Research Center for Space and Cosmic Evolution, Ehime University, 2-5 Bunkyo-cho, Matsuyama, Ehime, Japan, 790-8577\\
$^{3}$National Astronomical Observatory of Japan, 2-21-1 Osawa, Mitaka, Tokyo 181-8588, Japan\\
$^{4}$Academia Sinica Institute of Astronomy and Astrophysics, 11F of Astronomy-Mathematics Building, AS/NTU, No.1, Section 4, Roosevelt Road, Taipei 10617, Taiwan\\
}

\date{Accepted XXX. Received YYY; in original form ZZZ}

\pubyear{2023}

\begin{document}
\label{firstpage}
\pagerange{\pageref{firstpage}--\pageref{lastpage}}
\maketitle

\begin{abstract}
Many massive quiescent galaxies have been discovered at $z>2$ thanks 
to multi-wavelength deep and wide surveys, 
however, substantial deep near-infrared spectroscopic observations are needed to constrain their 
star-formation histories statistically.
Here, we present a new technique to select quiescent galaxies with a short 
quenching timescale ($\leq0.1$ Gyr) at $z\sim2$ photometrically.
We focus on a spectral break at $\sim1600$ \AA~that appears 
for such fast-quenching galaxies $\sim1$ Gyr after quenching 
when early A-type stars go out, but late A-type stars still live.
This spectral break at $z\sim2$ is similar to a Lyman break at $z\sim4$.
We construct a set of color criteria for $z\sim2$ fast-quenching galaxies on $g-r$ vs. $r-i$ and $i-J$ vs. $J-H$ 
or $\rm i-[3.6]$ vs. $\rm [3.6]-[4.5]$ color diagrams, 
which are available with the existing and/or future wide imaging surveys, 
by simulating various model galaxy spectra and test their robustnesses 
using the COSMOS2020 catalog.
Galaxies with photometric and/or spectroscopic redshifts $z\sim2$ 
and low specific star formation rates are successfully selected using these colors.
The number density of these fast-quenching galaxy candidates at $z\sim2$ suggests that
massive galaxies not so far above the star-formation main sequence at $z=3-4$ should be their progenitors. 
\end{abstract}




\section{Introduction}

Now many massive quiescent galaxies at $z>2$ 
have been confirmed spectroscopically by observing their rest-frame optical spectra 
using the near-infrared (NIR) spectrographs on the 10 m class 
ground-based telescopes or the James Webb Space Telescope (JWST)
(e.g., \citealt{2018A&A...618A..85S,2020ApJ...889...93V,2023ApJ...947...20V,2023MNRAS.520.3974C,2023Natur.619..716C}).
Such distant quiescent galaxies are characterized by low star formation rates (SFR) and significant Balmer breaks 
that are clues of the rapid evolution of massive galaxies in the early Universe. 

To understand how star formation in a galaxy occurred and was quenched, 
we need to characterize its star-formation history (SFH), 
which can be constrained based on the stellar spectral features like Balmer absorption in the rest-frame optical.
One of the key parameters of the SFH is the timescale of quenching star formation
which is critical to understanding how massive galaxies today have stopped star formation and started passive evolution. 
The timescales of quenching range widely from so-called fast-quenching 
(i.e., expressed in an exponentially decaying SFR with time $t$ as SFR $\propto\exp(-t/\tau)$ with $e$-folding time $\tau\sim0.1$ Gyr)
and slow-quenching galaxies  ($\tau\sim1$ Gyr) \citep{2019ApJ...874...17B}.
In the early Universe ($z>2$), environmental effects like ram-pressure stripping 
may not be major quenching mechanisms. 
The gas-rich major merger and violent disk instability inducing rapid gas exhaustion,
and sudden removal of gas via active galactic nuclei (AGN) feedback 
are plausible mechanisms for fast-quenching \citep{2015MNRAS.450.2327Z}.
Normal star-forming galaxies have gas depletion times of $\sim1$ Gyr (e.g., \citealt{2010MNRAS.407.2091G})
and thus, if the cold gas supply is halted for some reason, e.g., heating of the gas via
virial shock \citep{2003MNRAS.345..349B} or radio mode feedback 
of AGN \citep{2006MNRAS.365...11C}, 
or galaxies are stabilized against collapse \citep{2009ApJ...707..250M}, 
galaxies should quench slowly. 

According to the analyses of the spectra of massive galaxies in the local Universe, 
they likely formed stars in shorter timescales as the stellar masses 
and formation redshifts become larger (e.g., \citealt{2005ApJ...621..673T}). 
Based on the deep NIR spectroscopy of massive quiescent galaxies at $z>2$,
they are likely formed via short periods of intense starburst shutting down quickly
(e.g., \citealt{2020ApJ...889...93V,2023ApJ...947...20V,2021ApJ...919....6K}). 
However, it is hard to study the SFHs of galaxies at $z>2$ statistically 
because deep spectroscopic observations are necessary to constrain the SFHs robustly.
There are several attempts to classify the fast and slow-quenching population 
photometrically using the rest-frame $UVJ$ color diagram 
(e.g., \citealt{2019ApJ...874...17B, 2022A&A...666A.141M}).
But to construct the rest-frame $UVJ$ color diagram, 
spectroscopic redshifts or reliable photometric redshifts based 
on multi-wavelength photometry are necessary. 

Here, we introduce a new method to extract massive quiescent galaxies
formed through fast ($\leq0.1$ Gyr) quenching
by simple color criteria using the optical and NIR or mid-infrared (MIR) photometry 
available from the existing or near future wide imaging surveys.
This paper is organized as follows: in section 2, we introduce 
the color criteria to select massive quiescent galaxies at $z\sim2$ 
constructed by a stellar population synthesis modeling.
In section 3, we test this selection method using the multi-band photometric catalog in the COSMOS field.
In section 4, possible interlopers and the connection 
with massive star-forming galaxies (SFGs) at high redshift are discussed, 
and section 5 is the conclusion.
We adopt cosmological parameters $\Omega_{\rm m} =0.3$, 
$\Omega_{\rm \Lambda} =0.7$ and $H_0 =70$ km s$^{-1}$ Mpc$^{-1}$, 
and \citet{2003PASP..115..763C} Initial Mass Function (IMF) in the whole paper. 
Magnitudes are all expressed in the AB system. 

\begin{figure}
	\includegraphics[width=\columnwidth]{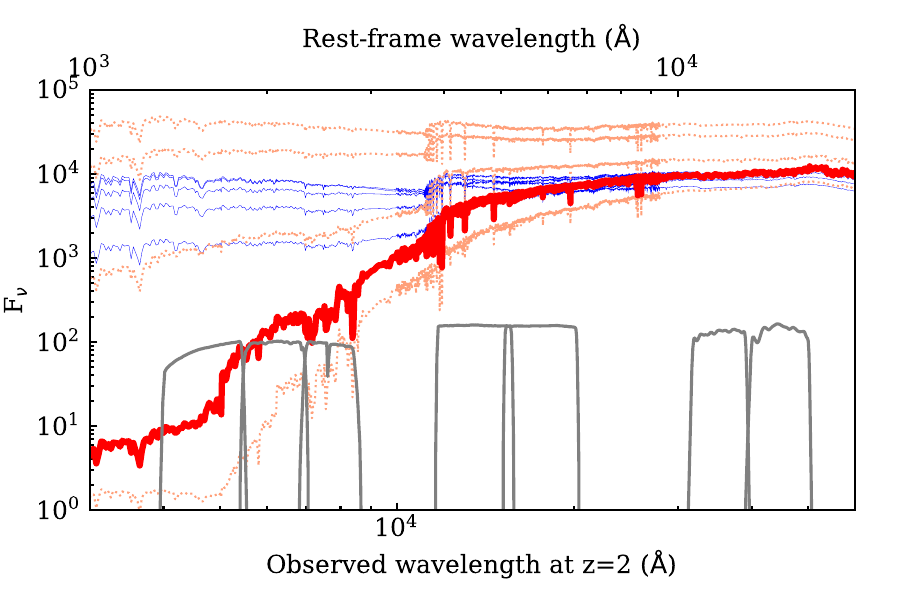}
    \caption{The evolution of model galaxy spectra with age. 
    Red and blue thin curves show the models with exponentially decay SFHs (SFR $\propto\exp(-t/\tau)$)
    with $\tau=0.1$ and 1 Gyr, respectively. 
    The models at the ages $t=$ 0.1, 0.2, 0.5, 1, and 2 Gyr are shown. 
    The thick red curve shows the model with $\tau=0.1$ Gyr and $t=1$ Gyr. 
    The gray curves show the transmission curves of the Subaru HSC $gri$, Euclid NISP $JH$, and $Spitzer$ IRAC [3.6][4.5].}
    \label{fig:sed_evol}
\end{figure}

\begin{figure}
	\includegraphics[width=\columnwidth]{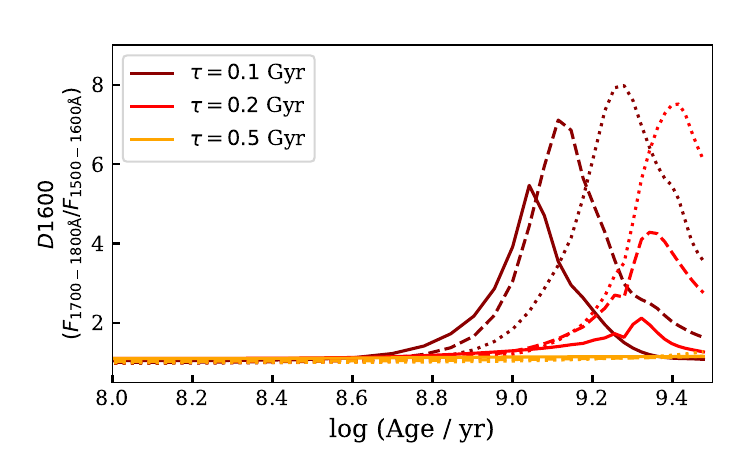}
    \caption{The dark red, red, and orange curves show the evolutions of the 1600 \AA~breaks (D1600)
    with age for exponentially decaying SFHs with $\tau=0.1$, 0.2, and 0.5 Gyr, respectively.
    The solid, dashed, dashed-dotted, and dotted curves show the models with $Z$ = 1, 0.4, \&~0.2 $Z_{\odot}$.}
    \label{fig:sed_evol2}
\end{figure}

\begin{figure*}
\centering
\includegraphics[width=170mm]{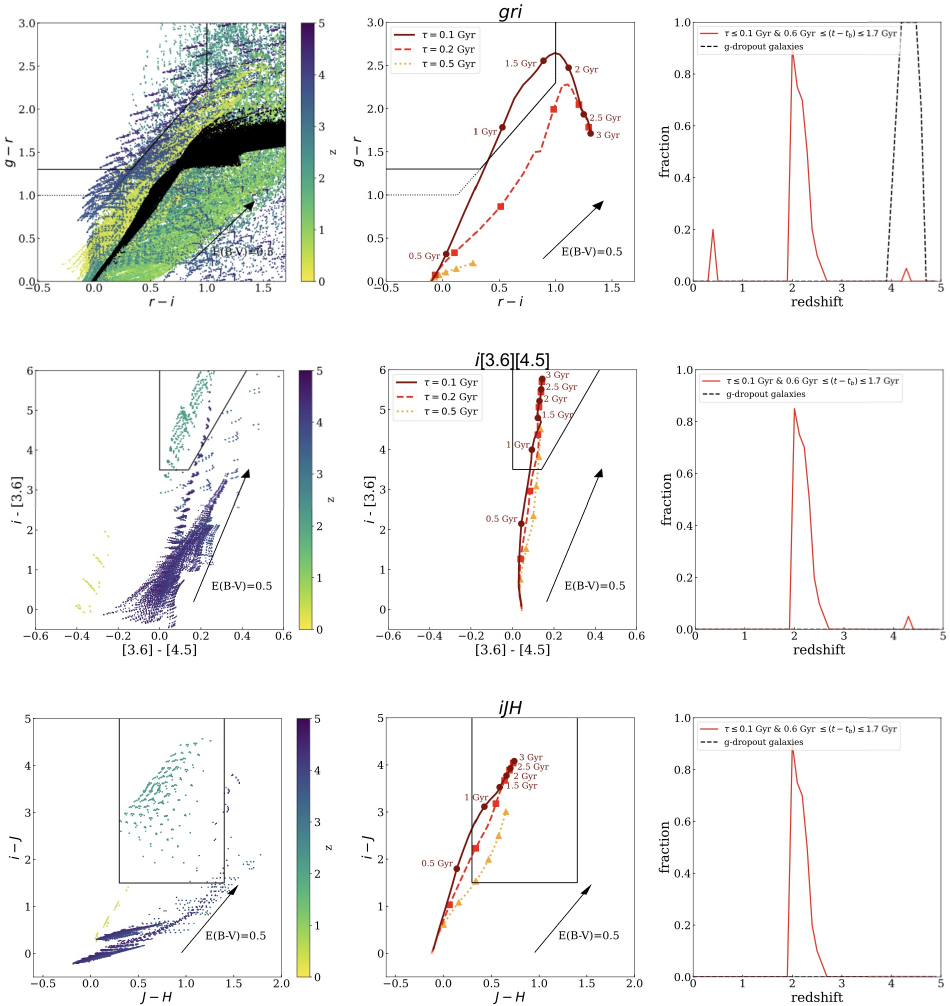}
    \caption{Top row: the left panel shows the model colors of galaxies on $g-r$ vs. $r-i$ color diagram. The colors of the points depend on the redshifts as shown in the color bar. The black stars show the dwarf stars from the TRILEGAL model \citep{2005A&A...436..895G}.
The black dotted line shows the color criterion for $g$-dropout LBGs \citep{2016ApJ...826..114T}.
The solid black line shows the color criterion to select $z\sim2$ quenched galaxies proposed in this study. The black arrow shows the reddening vector with $E(B-V) = 0.5$.
The central panel shows the evolution of colors with age. The dark red, red, and orange curves with points show the models with $t_{\rm burst}=0$ and $\tau=0.1, 0.2$ \& $0.5$ Gyr, respectively, and $E(B-V) = 0$ from $t=0-3$ Gyr. The points are shown with 0.5 Gyr steps.
The right panel shows the redshift selection function of $gri$ color criterion.
The red curve shows the selection function for fast quenching galaxies with $\tau\leq0.1$ Gyr and 0.6 $\leq (t-t_{\rm burst})$/Gyr $\leq$ 1.7. The black dashed curve shows that for star-forming galaxies ($\tau=10$ Gyr, 0.1 $\leq (t-t_{\rm burst})$/Gyr $\leq$ 0.5, $E(B-V)\leq0.1$).
Middle row: similar to the top row but for $i-$ [3.6] vs. [3.6] - [4.5] colors. The right panel shows the redshift selection function for models satisfying both $gri$ and $i$[3.6][4.5] color criteria.
Bottom row: similar to the top row but for $i-J$ vs. $J-H$ colors. The right panel shows the redshift selection function for models satisfying both $gri$ and $iJH$ color criteria.}
    \label{fig:color_model}
\end{figure*}

\section{New technique to select fast-quenching galaxies}

Figure \ref{fig:sed_evol} shows the evolution of the model spectral energy distributions (SEDs) 
of galaxies with ages $t=0.2-2$ Gyr for exponentially decay (SFR $\propto\exp (-t/\tau)$) SFHs with solar metallicity.
There is a significant spectral break at $\sim1600$ \AA~
for the model with  $\tau=$ 0.1 Gyr and age $\sim1$ Gyr when most B to early A-type stars go out 
but late A to F-type stars are still alive.
Figure \ref{fig:sed_evol2} shows the evolution of the  
ratio of the fluxes at $1200-1600$ \AA~and $1600-2000$ \AA~(named D1600), 
for exponentially decay models with $\tau=$ 0.1, 0.2, and 0.5 Gyr 
with metallicities $Z$ = 1, 0.4, \&~0.2 $Z_{\odot}$.
The smaller $\tau$ models show a sharper rise and fall in D1600.

We examined a selection technique for such fast-quenching galaxies
by detecting this 1600 \AA~break with existing and/or future deep and wide imaging surveys.
At $z\sim2$, it drops at $\sim5000$ \AA~or between $g$ and $r$-bands 
similar to Lyman break galaxies (LBGs) at $z\sim4$. 
We calculated the colors of galaxies at $z=0-4.8$ for $g,~r,$ \& $i$-bands 
of the Hyper Suprime-Cam (HSC) on Subaru Telescope \citep{2018PASJ...70S...1M},  
the [3.6] \& [4.5]-bands of the Infrared Array Camera (IRAC) on the {\it Spitzer} \citep{2004ApJS..154...10F}, 
and the $J$ \& $H$-bands of the Near Infrared Spectrometer and Photometer (NISP)  on the Euclid \citep{2022A&A...662A..92E} 
using {\sf CIGALE} \citep{2019A&A...622A.103B}.
In {\sf CIGALE}, galaxy models are generated by adopting 
 \citet{2003MNRAS.344.1000B} ({\sf BC03})
stellar population synthesis modeling code, \citet{2003PASP..115..763C} 
Initial Mass Function (IMF), and solar metallicity.
The time evolution of SFR is defined as, 
\begin{equation}
\rm     SFR(t) \propto 
     \begin{cases} {{A} \ {\rm for} \ t  <t_{\rm burst},}\\
     				{A\times\exp(-(t-t_{\rm burst})/\tau) \ {\rm for} \ t \geq t_{\rm burst},}
    \end{cases}
\end{equation}
where $A$ is constant, $t_{\rm burst}$ is the time duration of the constant star formation
and $\tau$ is the $e$-folding time. 
 Hereafter, the characteristic properties of an SFH are defined following \citet{2018A&A...618A..85S}; 
the formation time $t_{\rm form}$ as the time when 50 \% of the total integrated star formation occurred, 
the quenching time $t_{\rm quench}$ as the time when the SFR 
is less than 10 \% of the $\rm \langle SFR\rangle_{\rm main}$  
where $\rm \langle SFR\rangle_{\rm main}$ is the mean SFR during 
the main stellar mass assembly episode enclosing $t_{\rm form}$ and 68\% of the total integrated SFR. 
The nebular emission line is added by adopting the templates based on \citet{2011MNRAS.415.2920I} parameterized with ionization parameter $U$, Lyman continuum escape fraction $f_{\rm esc}$ and partial absorption by dust before ionization $f_{\rm dust}$, and metallicity $Z$ 
which is assumed to be the same as a stellar one.
The dust attenuation is added by adopting the \citet{2000ApJ...533..682C} extinction law with a constant continuum to line $E(B-V)$ ratio, $E(B-V)_{\rm continuum}/E(B-V)_{\rm line} = 0.44$.
Table \ref{tab:params} summarizes the ranges of the parameters.  

\begin{table}
	\centering
	\caption{Model parameters}
	\label{tab:params}
	\begin{tabular}{lcc} 
		\hline
		& unit & values \\
		\hline
$t_{\rm burst}$		& Gyr & 0.1, 0.2 \& 0.5\\
$\tau$		& Gyr & 0.05,~0.1,~0.2,~0.5,~1,~2,~5, \&~10 \\
$t$ & Gyr & 0.1 to 3.0 with 0.1 steps, 5, 10 \\
$\frac{E(B-V)_{\rm continuum}}{E(B-V)_{\rm line}} $ & - & 0.44\\
$E(B-V)_{\rm lines} $ & mag & 0.0 to 2.0 with 0.4 steps\\
$\log U$ & - & -1, -2  \& -3 \\
$f_{\rm esc}$ & - & 0.3, 0.6 \& 1.0 \\
$f_{\rm dust}$ & - & 0.0\\
line width & km s$^{-1}$ & 200 \\
		\hline
	\end{tabular}
\end{table}
 
Figure \ref{fig:color_model} top row shows the model colors in $g-r$ vs. $r-i$. 
The black dotted line shows the color criterion 
for LBGs at $z\sim4$ \citep{2016ApJ...826..114T}, which works well to remove low-z objects.
In addition to LBGs at $z\sim4$, namely, some galaxy models with $\tau\leq0.1$ Gyr at $z\sim0.5$ or $z\sim2$
 fall within this color criterion.
The central panel shows the color evolution track for models 
with $\tau=0.1, 0.2$ \& 0.5 Gyr, $t_{\rm burst}=0.0$ Gyr, and $E(B-V)_{\rm lines}=0.0$ without emission line. 
There is no significant difference in the models with different $t_{\rm burst}$.
At $z=1.9-2.3$, galaxies with $\tau\leq0.1$ Gyr
and the $t_{\rm form} - t_{\rm quench}\lesssim0.3$ Gyr 
pass through the $gri$-color criterion when they are $0.7\sim1.6$ Gyr after $t_{\rm burst}$
or $0.5\sim1$ Gyr after $t_{\rm quench}$.

Figure \ref{fig:color_model} middle and bottom rows show the $i - $[3.6] vs. [3.6] $-$ [4.5] 
and  $i - J$ vs. $J-H$ color diagrams
for the galaxies satisfying the $g$-dropout color criterion slightly modified to select $z\sim2$ galaxies preferentially (Eq. \ref{eq:color_criteria1} in the following, solid line in Figure \ref{fig:color_model} top).
The combinations of $WISE$ and $JWST$ filters are presented in the Appendix \ref{sec:diff_filter}. 
The $g$-dropout galaxies at different redshifts are well separated on both color diagrams.
Then, galaxies rapidly quenched ($\tau\leq0.1$ Gyr) and $\sim1$ Gyr after a quenching at $z\sim2$
can be selected by the combination of the first color criterion, 
\begin{equation}
     \begin{cases} {g-r > 1.3,\ -1<r-i<1,}\\
     				{g-r>1.5\times(r-i)+0.8,}\\
    \end{cases}\label{eq:color_criteria1}
\end{equation}
hereafter $gri$ color criterion, 
and the second color criterion, 
\begin{equation}
     \begin{cases} 
          			{i-\rm[3.6] >3.5,\ [3.6]-[4.5]>0,}\\
				{i-\rm[3.6]> 8.0 \times ([3.6]-[4.5])+2.5,}
    \end{cases}\label{eq:color_criteria2}
\end{equation}
hereafter the $i$[3.6][4.5] color criterion, or
\begin{equation}
     i-\rm J >1.5,\ 0.3<J-H<1.4,
     \label{eq:color_criteria3}
\end{equation}
hereafter the $iJH$ color criterion. 

The color criteria are set also considering the colors of the observed galaxies 
at photometric and spectroscopic redshifts $z\sim2$ (see section \ref{sec:results})
not to miss a large fraction of them.
The right panel of Figure \ref{fig:color_model} shows the redshift selection functions.
The red curve shows that for fast-quenching galaxies
selected with the above color criteria calculated by assuming models with $\tau\leq 0.1$ Gyr,  $0.7 \leq (t-t_{\rm burst})$/Gyr $\leq 1.6$ and $E(B-V)=0$.
The black dashed curve shows that for star-forming galaxies like LBGs calculated by assuming models with $\tau=10$ Gyr, 0.1 $\leq (t-t_{\rm burst})$/Gyr $\leq$ 0.5, $E(B-V)\leq0.1$. 
The latter are removed after applying the second color criterion.
The metallicity dependence of the galaxy colors is discussed in Appendix \ref{sec:diff_metal}. 

\begin{figure*}
	\includegraphics[width=170mm]{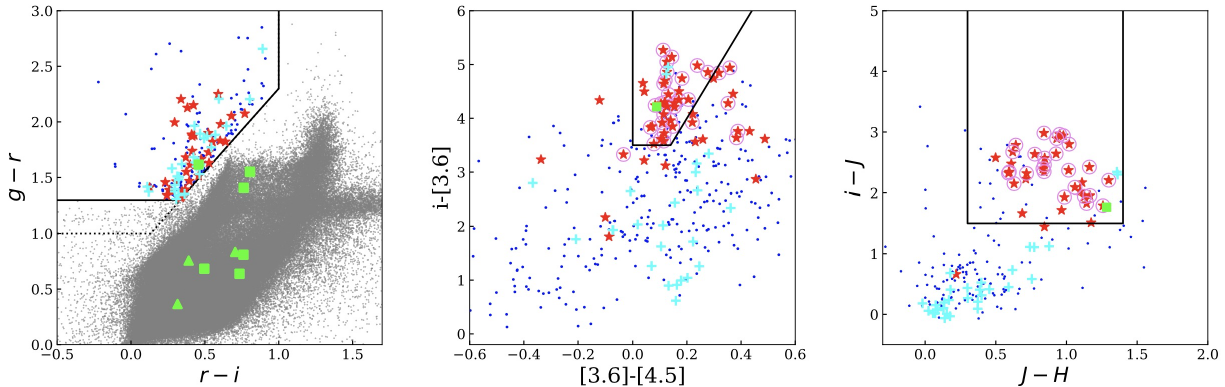}
    \caption{Left: The $g-r$ vs. $r-i$ color diagram for galaxies in the COSMOS2020 catalog. 
    The small blue dots show the objects satisfying the $gri$ color criterion. 
    The red stars and cyan crosses show those with $1.9<z_{\rm phot}<2.3$ 
    and with $1.9<z_{\rm spec}<2.3$ (HSC-SSP PDR3 spectroscopic redshift catalog 
    \citealt{2022PASJ...74..247A}), respectively. 
    The green-filled triangles and squares show massive quiescent galaxies 
    with $1.9<z_{\rm spec}<2.3$ 
    in \citet{2017ApJ...834...18B} and \citet{2020ApJ...888....4S}, respectively.
    The gray dots show the distribution of the whole galaxies.
    Center: The $i-$ [3.6] vs. [3.6] $-$ [4.5] color diagram for objects satisfying Eq. \ref{eq:color_criteria1}. The solid black line shows the Eq. \ref{eq:color_criteria2} color criterion. The circled objects also satisfy the $iJH$ color criterion. 
    Right: Similar to the center but for $i-J$ vs. $J-H$ and Eq. \ref{eq:color_criteria3}. The circled objects also satisfy the $i$[3.6][4.5] color criterion.}
    \label{fig:cosmos_color_z2}
\end{figure*}

\begin{figure*}
	\includegraphics[width=80mm]{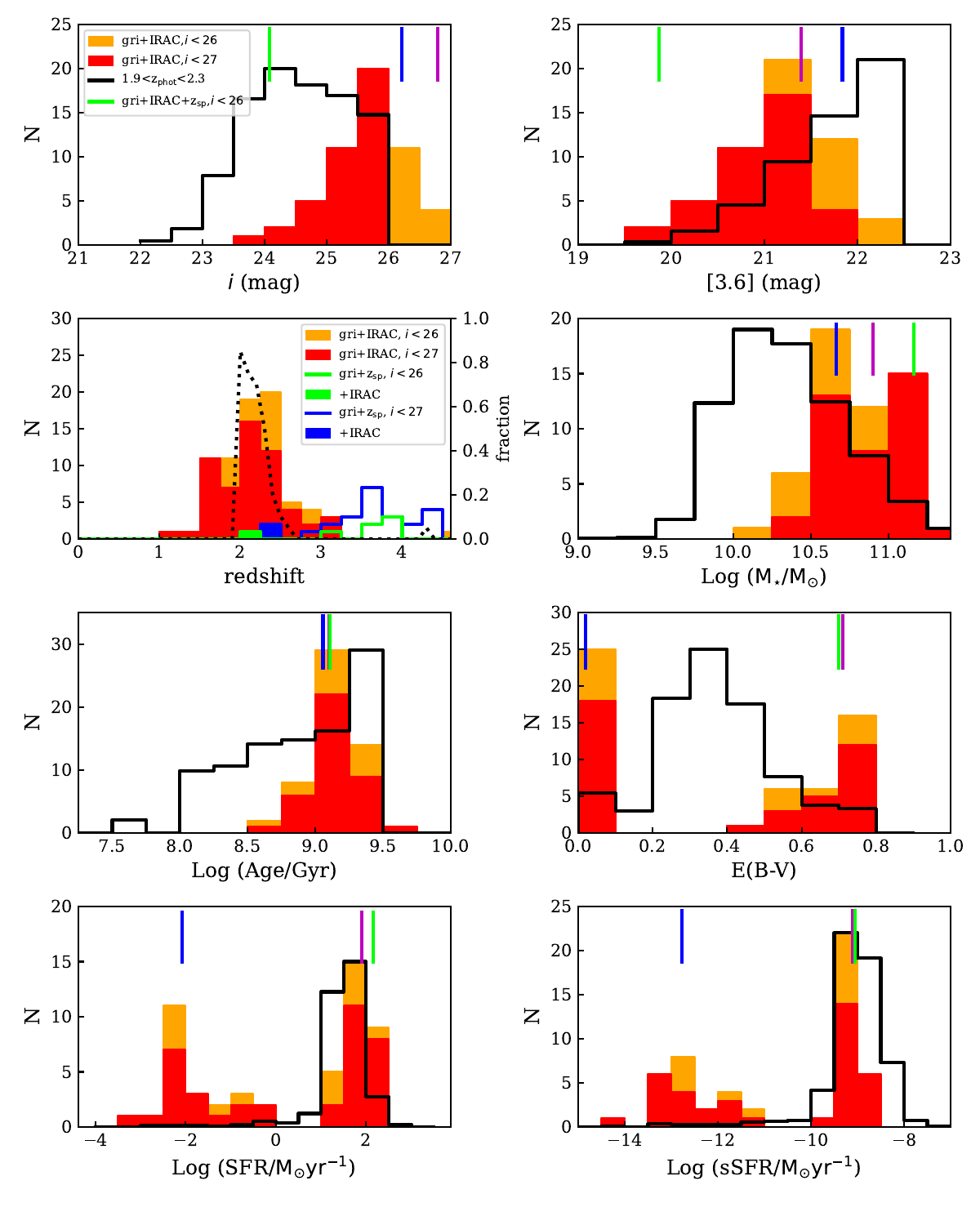}
	\includegraphics[width=80mm]{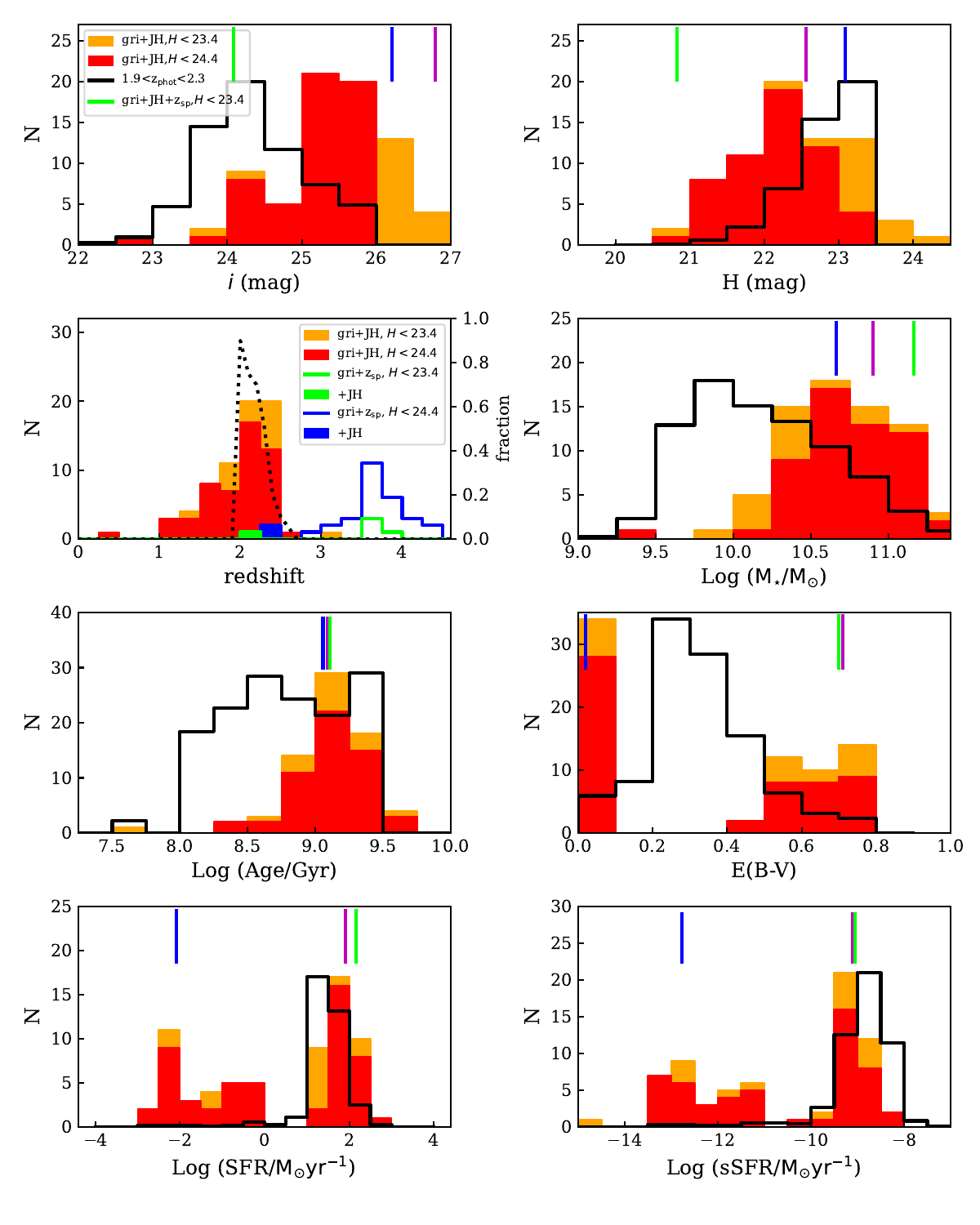}
    \caption{Left: The distributions of $i$ and [3.6] magnitudes, photometric redshifts, 
   stellar masses, ages, $E(B-V)$, SFRs, and sSFRs from top left to bottom right for galaxies selected by the $i$[3.6][4.5] color criterion.
   The red and orange filled histograms show the objects selected 
   with our color criteria with $i<26$ and [3.6] $<22.5$, and $i<27$ and [3.6] $<23.5$, respectively.
   The black line histograms show the normalized distribution 
   of the galaxies with $i<26$ and [3.6] $<22.5$ and $1.9<z_{\rm phot}<2.3$.
   The blue, magenta, and green vertical lines show 3D-HST 11142, 
   3D-HST 16527, and UV-230929 in Table \ref{tab:specz}, respectively.
   On the redshift distribution diagram, the blue and green histograms show the spectroscopic redshift distribution for galaxies satisfying $gri$ color criterion with $i<26$ and [3.6] $<22.5$, and $i<27$ and [3.6] $<23.5$, respectively, 
   while those satisfying the $i$[3.6][4.5] color criterion are shown with filled histograms. 
   The black dotted curve shows the redshift selection function presented in Figure \ref{fig:color_model}.
   Right: Similar to the left panel but for the $iJH$ color criterion. 
}
    \label{fig:cosmos_sedparams}
\end{figure*}

\begin{figure*}
	\includegraphics[width=120mm]{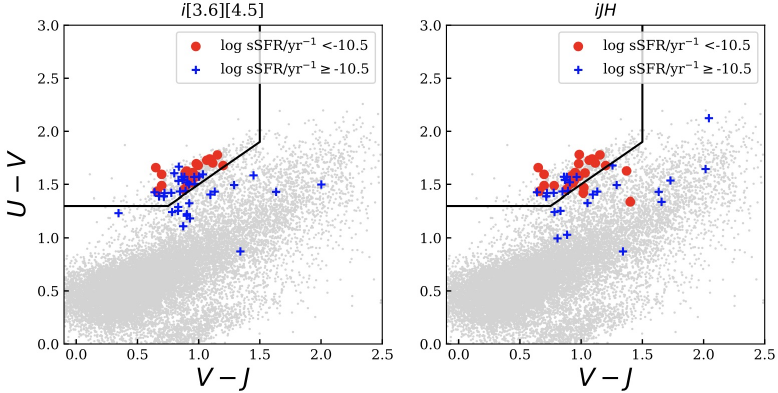}
    \caption{Left: The rest-frame $UVJ$ color diagram. 
    The rest-frame $UVJ$ colors are interpolated from the best-fit SED models. 
    The red circles and blue crosses show the galaxies selected 
    by the IRAC $i$[3.6][4.5] color criterion and with $\log$ (sSFR/M$_{\odot}$ yr$^{-1}) <-10.5$ and $\geq -10.5$ yr$^{-1}$, respectively.
    The gray dots show the galaxies with $1.9<z_{\rm phot}<2.3$ and $H<23.4$.
    The black line shows the color criterion for quiescent galaxies by \citet{2012ApJ...745..179W}.
    Right: Similar to the left panel but for the $iJH$ color criterion.
}
    \label{fig:uvj}
\end{figure*}

\section{Result}
\label{sec:results}

We tested the above color criteria using the COSMOS2020 catalog \citep{2022ApJS..258...11W}.
1.279 deg$^2$ area is observed with the HSC, UltraVISTA, and IRAC after masking the bright star halos \citep{2022arXiv221202512W}.
From the CLASSIC catalog, we used the 3 arcsec aperture photometry of the HSC $g,$ $r,$ \& $i$
and the UltraVISTA $J~\&~H$
and the PSF fitted photometry for the IRAC [3.6] 
and [4.5], whose 5 $\sigma$ detection limits are 
27.5, 27.2, 27.0, 24.7, 24.4, 25.7, and 25.6 mag, respectively.
The UltraVISTA $JH$ responses differ from the Euclid NISP $JH$ but do not change the color criterion critically.
The photometric redshifts ($z_{\rm phot}$), and the best-fit ages, SFRs, specific SFRs (sSFR), stellar masses, 
and $E(B-V)$ are those measured 
with {\sf Le Phare} \citep{2002MNRAS.329..355A,2006A&A...457..841I} 
using the $GALEX$ NUV to IRAC [4.5] $\mu$m-bands photometry.
The ages from {\sf Le Phare} are the time from galaxy formation.
The accuracy of the photometric redshifts of the catalog is $\sigma < 0.025 (1+z)$ at $i<25$ in typical.
The color criteria were tested for galaxies with $i<26$ and [3.6] $<22.5$ mag or $i<26$ and $H<23.4$ mag.

\subsection{i[3.6][4.5] color criterion}

Figure \ref{fig:cosmos_color_z2} left and center shows 
the $g-r$ vs. $r-i$ and $i-$ [3.6] vs. [3.6] $-$ [4.5] color diagrams. 
The galaxies satisfying the $gri$ color criterion
and those with $1.9<z_{\rm phot}<2.3$ are presented 
in the $i-$ [3.6] vs. [3.6] - [4.5] color diagram (center). 
The massive quiescent galaxies confirmed at $1.9<z_{\rm spec}<2.3$ 
detecting the Balmer break and absorptions spectroscopically 
in \citet{2017ApJ...834...18B} and \citet{2020ApJ...888....4S}, 
and the galaxies from the spectroscopic redshift catalog of the COSMOS field summarized by HSC-SSP PDR3 \citep{2022PASJ...74..247A} are also presented.

57 out of 359 galaxies satisfying the $gri$ color criterion also satisfy the $i$[3.6][4.5] color criterion.
Three galaxies with spectroscopic redshifts satisfy the $i$[3.6][4.5] color criterion (section \ref{sec:specz}).
Figure \ref{fig:cosmos_sedparams} shows the distribution of $i$ and [3.6] magnitudes, 
photometric redshifts, and SED parameters of the galaxies selected 
by the $i$[3.6][4.5] color criterion compared to those of galaxies with $1.9<z_{\rm phot}<2.3$.
The $i$[3.6][4.5] color criterion selects galaxies in a skew redshift range around $z\sim2$.
Relatively massive galaxies are selected compared to the galaxies at $1.9<z_{\rm phot}<2.3$.
There is a significant spike at age $\sim1$ Gyr.
There are two peaks in each SFR and specific SFR (sSFR) distribution. 
22 out of the 57 galaxies have $\log $ (sSFR /M$_{\odot}$ yr$^{-1}) <-10.5$ 
or regarded as quiescent galaxies in the best-fit SEDs.
The color criterion works similarly for fainter galaxies with $i<27$ and [3.6] $<23.5$.
Figure \ref{fig:uvj} shows the rest-frame $UVJ$ color diagram 
often used to classify the quiescent galaxies (e.g., \citealt{2009ApJ...691.1879W}).
Most of the galaxies selected by the $i$[3.6][4.5] color criterion also satisfy 
or around the $UVJ$ color criterion for quiescent galaxies 
though some objects with $\log $ (sSFR /M$_{\odot}$ yr$^{-1}) \geq-10.5$ are outliers.

Since SED fitting can suffer from the degeneracy of dusty SFG and quiescent galaxy models, 
we investigated the {\it Spitzer} Multi-Band Imaging Photometer (MIPS) 24 $\mu$m to limit the obscured SFR.
First, we checked the individual detections 
on the 24 $\mu$m image, which has a 5$\sigma$ detection limit of 70 $\mu$Jy.
We selected the galaxies within 3 arcsec of the sources listed 
in the S-COSMOS \citep{2007ApJS..172...86S} G03 catalog
and checked the images of them by eye.
The 0/20 and 2/25 galaxies with $\log$ (sSFR / yr$^{-1}) <-10.5$ and $\geq -10.5$, respectively,
in the 24 $\mu$m FoV are detected individually on the 24 $\mu$m. 
The objects detected with 24 $\mu$m have fluxes of $195\pm14$ $\mu$Jy \& $348\pm16$ $\mu$Jy.
Adopting the median of the empirical SED models in \citet{2017ApJ...840...78D}, 
they correspond to SFRs of $\sim1000$ and $2000$ M$_{\odot}$ yr$^{-1}$ though AGNs (e.g., \citealt{2015PASJ...67...86T}) 
can also contribute to their 24 $\mu$m emission.
After rejecting the sources detected on 24 $\mu$m and/or close to the bright sources,
a median stacking analysis of the sample was performed.
The flux value and error are the median and standard deviation of 1000 times boot-strap resampling.
Table \ref{tab:ir_flux} lists the result of the stacking analysis for all, 
$\log$ (sSFR / yr$^{-1}) <-10.5$, and $\geq -10.5$, where there is no detection above 2 $\sigma$ for all subsample.
The 2 $\sigma$ upper limits on the 24 $\mu$m fluxes of the all, $\log$ (sSFR /yr$^{-1}) <-10.5$, 
and $\log$ (sSFR / yr$^{-1}) \geq-10.5$ galaxies correspond to the SFRs of $<51$, $<51$ and  $<81$ M$_{\odot}$ yr$^{-1}$, respectively.
These upper limits are lower than the SFR at the main sequence 
(SFR$_{\rm MS}\approx87$ M$_{\odot}$ yr$^{-1}$; \citealt{2015A&A...575A..74S})
for the median stellar mass of the sample ($\log (M_{\star}/$M$_{\odot}) \approx 10.7$) at $z=2$.

Eq. \ref{eq:color_criteria1} is enclosed in the $g$-dropout color criterion 
for LBGs at $z\sim4$ \citep{2016ApJ...826..114T}.
$0.5$ (7)\% of the $g$-dropout galaxies with $i<26$ ($i<26$ and [3.6] $ <22.5$) 
in the COSMOS2020 catalog satisfy our color criteria. 
Thus the contamination of $z\sim2$ galaxies in $g$-dropout galaxies is negligible, 
though it can not be negligible for those also bright in the MIR/NIR. 

\begin{table}
	\centering
	\caption{The 2$\sigma$ upper limits on the 24 $\mu$m fluxes of the candidate fast-quenching galaxies derived from median stacking.}
	\label{tab:ir_flux}
	\begin{tabular}{llcc} 
		\hline
							& 			& $N_{\rm stack}$ & $F_{\rm 24 \mu m}$ \\
							& 			& & ($\mu$Jy) \\
		\hline
$gri+$ [3.6] \& [4.5]		& All               & 41 & $<10.2$   \\
($[3.6] <22.5$ \& $i<26$)	& Low sSFR  & 17 & $<10.2$ 	\\
							& High sSFR & 23 & $<15.9$\\
$gri+JH$	       & All               & 39 & $<10.5$   \\
($H <23.4$ \& $i<26$	)	& Low sSFR  & 21 & $<11.2$ \\
							& High sSFR & 18 & $<23.5$	\\
		\hline
	\end{tabular}
\end{table}

\subsection{iJH color criterion}

Figure \ref{fig:cosmos_color_z2} right 
show $i-J$ vs. $J-H$ color diagram.
53 galaxies are selected by applying the $gri$ and $iJH$ color criterion, 
where 373 galaxies satisfy Eq. \ref{eq:color_criteria1}.
The SED parameters of the galaxies selected with the $iJH$ color criterion 
are similar to those selected using the $i$[3.6][4.5] color criterion. 
All the spectroscopically confirmed galaxies selected with the $i$[3.6][4.5] color criterion
are also selected by the $iJH$ color criterion.
The 2/27 and 4/22 galaxies with $\log$ (sSFR/M$_{\odot}$ yr$^{-1}) <-10.5$ 
and $\geq -10.5$ yr$^{-1}$, respectively, 
in the 24 $\mu$m FoV are detected individually on the 24 $\mu$m. 
Similarly, there is no detection above $2\sigma$ significance for their stacking 
after rejecting these individually detected sources (Table \ref{tab:ir_flux}). 
We compare the results of the $JH$ and $i$[3.6][4.5] color criteria in section \ref{sec:dis_jh_irac}.

\subsection{Spectroscopically confirmed galaxies}
\label{sec:specz}

One object from \citet{2020ApJ...888....4S} (UV-230929) 
and two objects from the 3D-HST (
\citealt{2012ApJS..200...13B, 2016ApJS..225...27M}, included in the HSC-SSP PDR3 spectroscopic redshift catalog)
are selected with both the $JH$ and $i$[3.6][4.5] color criteria. 
Their photometric properties from the COSMOS2020 catalog 
are presented in Figure \ref{fig:cosmos_sedparams} and Table \ref{tab:specz}.
Our color criteria extract only one object from massive quiescent galaxies 
confirmed spectroscopically at $z\sim2$ in \citet{2017ApJ...834...18B} and \citet{2020ApJ...888....4S}
because our color criteria are only sensitive to massive quiescent galaxies 
with specific quenching timescales and ages.

\begin{figure}
	\includegraphics[width=80mm]{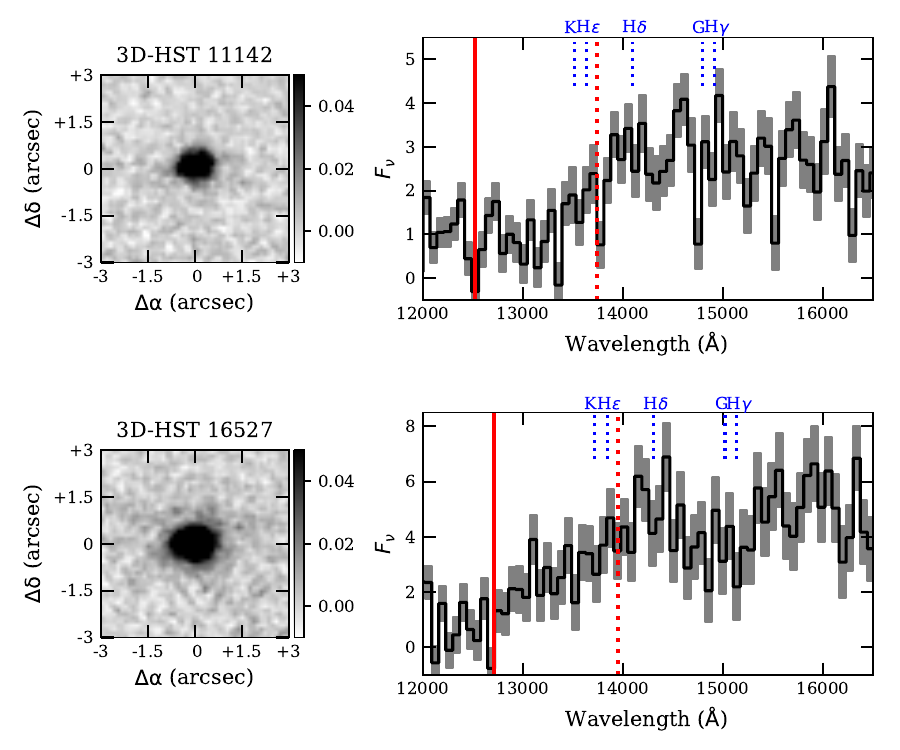}
    \caption{The $HST$ $F160W$-images of 3D-HST 11142 and 3D-HST 165279 from CANDELS \citep{2011ApJS..197...35G,2011ApJS..197...36K}.
    The spectra (black) and errors (gray) are from 3D-HST survey \citep{2012ApJS..200...13B, 2016ApJS..225...27M}.
    The blue vertical lines show the expected locations for H$\epsilon$, H$\delta$, H$\gamma$, and GHK absorption lines.
    The red solid and dashed vertical lines show that for Balmer and 4000~\AA~breaks, respectively.
}
    \label{fig:speczimage}
\end{figure}

Based on the spectroscopic analyses in \citet{2020ApJ...888....4S}, 
UV-230929 has a stellar mass of $\log (M_{\star}/ $M$_{\odot})=11.48^{+0.16}_{-0.16}$, 
$\log\rm(Age/yr) = 9.10^{+0.28}_{-0.28}$
and the upper limit of SFR from [O {\footnotesize II}] $<4$ M$_{\odot}\rm yr^{-1}$.
The spectra of 3D-HST 11142 and 16527 are presented in Figure \ref{fig:speczimage}.
They show Dn(4000) $\sim1.4$, consistent with a single burst model with age $\sim1$ Gyr \citep{1999ApJ...527...54B}.
There is no emission and absorption line above 2$\sigma$ significance. 
All of them are not detected individually in 24 $\mu$m.
Two are in the existing ALMA data (PID \#2015.1.00379.S \&~\#2013.1.01292.S) 
and not detected with $>2\sigma$ significance. 
Their SFR upper limits based on the ALMA are $\lesssim 100$ M$_{\odot}$ yr$^{-1}$, 
similarly assuming the median SED model of \citet{2017ApJ...840...78D}. 
From these above, all the spectroscopically confirmed objects satisfying our color criteria 
are well characterized as massive quiescent galaxies at $z\sim2$ 
though two are misidentified as dusty SFGs in the photometric best-fit SEDs.

They all have the $F160W$-band images taken 
with the Hubble Space Telescope (HST) Wide Field Camera 3 (WFC3).
Table \ref{tab:specz} presents the rest-frame optical half-light semi-major axes ($r$), S\'ersic indices ($n$), 
axis ratios measured on $F160W$ images \citep{2012ApJS..203...24V,2020ApJ...888....4S}.
Figure \ref{fig:speczimage} shows the images of 3D-HST 11142 and 16527 
from Cosmic Assembly Near-Infrared Deep Extragalactic Legacy Survey 
(CANDELS, \citealt{2011ApJS..197...35G,2011ApJS..197...36K}).
They are fitted by compact ($r_e\sim1$ kpc) and elliptical ($n>2.5$) morphologies. 
Their sizes are similar to those of quiescent galaxies with similar stellar masses at $z\sim2$ \citep{2014ApJ...788...28V}.

\begin{table*}
	\centering
	\caption{The spectroscopically confirmed galaxies satisfying the color criteria.}
	\label{tab:specz}
	\begin{tabular}{lcccccc} 
		\hline
ID		& $z_{\rm spec}$ &  $\log(M_{\star}/$M$_{\odot})$ & $r_{\rm maj}$ (kpc) &  $n$ &  axis ratio &  ALMA \\
		\hline
3D-HST 11142		& 2.48685 & $10.70^{+0.04}_{-0.04}$ &  $1.30\pm0.14$ & $8.0\pm1.6$ & $0.33\pm0.06$ & ...\\
3D-HST 16527		& 2.43494 & $10.96^{+0.06}_{-0.06}$ & $1.27\pm0.09$ & $4.0\pm0.6$ & $0.55\pm0.04$ & $F_{\rm 1.25 mm}<0.11$ mJy (2$\sigma$)\\
UV-230929		& 2.1679 & $11.48^{+0.16}_{-0.15}$ & $1.74\pm0.17$ & 3.01 & 0.73 & $F_{\rm 0.87 mm}<0.36$ mJy (2$\sigma$)\\
		\hline
	\end{tabular}
\end{table*}

\section{Discussion}
\label{sec:discussion}

\subsection{iJH vs. i[3.6][4.5] color criterion}
\label{sec:dis_jh_irac}

The objects satisfying both color criteria are circled in Figure \ref{fig:cosmos_color_z2}.
Among the 34 (26) objects with $i<26$, [3.6] $<22.5$, 
and $H<23.4$ satisfying the $iJH$ ($i$[3.6][4.5]) color criterion, 
26 (26) objects also satisfy the $i$[3.6][4.5] ($iJH$) color criterion.
Thus, the objects selected with the $i$[3.6][4.5] color criterion are not missed by the $iJH$ color criterion. 
Most of the objects selected with the $iJH$ color criterion
but missed by the $i$[3.6][4.5] color criterion have the $i-$ [3.6] vs. [3.6] $-$ [4.5] 
colors expected for dusty SFGs at $z\sim3$. 
It is because the $iJH$ color criterion is set not to miss a large fraction 
of model and observed galaxies, especially spectroscopically confirmed ones, 
at $z\sim2$, but encloses dusty SFG models at $z\sim3$. 
More outliers are on the $UVJ$ color diagram than those selected with the $i$[3.6][4.5] color criterion (Figure \ref{fig:uvj}).
Furthermore, $JH$-bands have more chances to catch notable emission lines like H$\alpha$ and [O {\footnotesize III}] though it is hard to consider the contribution of such emission lines robustly.
From these above, the galaxies selected with the $iJH$ color criterion
are likely more contaminated by dusty SFGs than those selected with the $i$[3.6][4.5] color criterion.

[3.6] and [4.5] are similar to $W1$ and $W2$ of the Wide-field Infrared Survey Explorer ($WISE$).
The unWISE reaches 50\% detection completeness at $W1 = 20.72$ mag in AB \citep{2019ApJS..240...30S}.
Thus we can survey the rare, most massive quiescent galaxies by combining the HSC-SSP or Legacy Survey of Space and Time (LSST) of the Rubin Observatory and $WISE$.
Several wide imaging surveys using the IRAC are also useful. 
We can perform a similar $JH$ selection using the Euclid and Roman space telescope surveys.
Quiescent galaxies with lower stellar masses can be selected, but large contamination from dusty SFGs is expected.
The calibration using the IRAC data in the Euclid Deep layer 
will be feasible to study massive quiescent galaxy populations effectively in the Euclid Wide layer.

\subsection{Completeness and purity of the selection method}

From these above, our color criteria extract  
massive quiescent galaxies at $z\sim2$ successfully.
These selection methods are only sensitive 
to massive quiescent galaxies quenched with $\tau\leq0.1$ Gyr 
and $\sim1$ Gyr after quenching at $z\sim2$.
Thus, it cannot completely select the whole quiescent galaxy population at $z\sim2$, 
as shown in Figure \ref{fig:cosmos_color_z2}.
Besides this incompleteness, it is still useful to survey fast-quenching galaxies.

The interlopers at low and high redshift are expected to be small according 
to the photometric redshift distribution presented in Figure \ref{fig:cosmos_sedparams}.
On the other hand, the contamination of dusty SFGs at $z\sim3$ is possible.
The $iJH$ color criterion is severely ($\sim$ one quarter at least) contaminated by dusty SFGs. 
Furthermore, more than half of galaxies satisfying our color criteria 
have sSFRs $>10^{-10}$ M$_{\odot}$ yr$^{-1}$ based on the SED fitting.
The discrepancy between the expected parameters from the color diagram and best-fit SEDs 
can be due to the shortage of the model range simulated in this study,
the degeneracy of the SED models, or the true contamination of dusty SFGs due to photometric errors.
First, the SED models adopted in \citet{2022ApJS..258...11W} for {\sf Le Phare} 
are the SED templates based on {\sf BC03} and the empirical SEDs of local Spirals/Ellipticals.
The {\sf BC03} templates in \citet{2022ApJS..258...11W} are simulated in Figure \ref{fig:color_model}.
By simulating the colors of local Spirals/Ellipticals based on observations with dust attenuation, 
we confirmed that these templates do not satisfy our color criteria.
Therefore, we conclude that the shortage in the SED modeling is not likely the cause of the high sSFR solution in the catalog.
The true contamination rate of dusty SFGs is not available with the existing data,
but given the upper limit on the sSFR based on the 24 $\mu$m, at least, 
our color criteria plausibly select the galaxies under the star-formation main sequence.

\subsection{The number density of fast-quenching galaxies}

There are 56 galaxies with $M_{\star}>10^{10.5}$ M$_{\odot}$ satisfying the $i$[3.6][4.5] color criterion. 
Conservatively, 22 galaxies with $M_{\star}>10^{10.5}$ M$_{\odot}$ satisfy the $i$[3.6][4.5] color criterion 
and have the $\log (sSFR/yr^{-1})<-10.5$ from the SED fitting.
Adopting the volume for the survey area of 1.279 deg$^2$ and redshift range of $1.9<z<2.3$,
the comoving number densities of galaxies with $M_{\star}>10^{10.5}$ M$_{\odot}$ 
selected by the IRAC criterion is $10^{-5.2}~(10^{-5.5}$ for conservative) cMpc$^{-3}$.

According to the models, the galaxies selected by our method should 
quench at $0.7-1.5$ Gyr before $z=2$, i.e., at $z=3-4$.
Thus, we compare the number density of fast-quenching galaxies 
with those of SFGs at $z=3-4$ to discuss 
the plausible progenitors of fast quenching galaxies at $z\sim2$.
SMGs are plausible progenitors of massive quiescent galaxies at $z>2$ 
because of the short depletion time (e.g., \citealt{2014PhR...541...45C}) in concordance 
with the bursty SFHs of massive quiescent galaxies at $z>2$.
Based on the SCUBA-2 Cosmology Legacy Survey for 2 deg$^2$, 
the number density of SMGs with SFR $>300$ M$_{\odot}$ yr$^{-1}$ 
which are mostly $M_{\star}>10^{10.5}$ M$_{\odot}$ at $z = 3 - 4$
is estimated as $6.42\pm0.70\times10^{-6}$ Mpc$^{-3}$ \citep{2017MNRAS.469..492M}.
It is comparable to or higher than the number density of massive quiescent galaxies at $z\sim2$ selected by our color criteria.

The number density of galaxies with $M_{\star}>10^{10.5}$ M$_{\odot}$ at $z=3-4.5$
is $10^{-3.8}$ cMpc$^{-3}$ by integrating the stellar mass function of galaxies 
in \citet{2022arXiv221202512W} measured based on the COSMOS2020 catalog.
Thus if 4 (2) \% of the galaxies with $M_{\star}>10^{10.5}$ M$_{\odot}$ at $z=3-4.5$ are immediately quenched,
the number density of the fast-quenching galaxies at $z=2$ can be explained. 
Counting the galaxies from high SFR/SFR$_{\rm MS}$ ones 
with $M_{\star}>10^{10.5}$ M$_{\odot}$ at $z_{\rm phot}=3-4.5$ using the COSMOS2020 catalog, 
the galaxies with above 1.5 (1.8 conservatively) $\times$ 
the star-formation main sequence at $z=3-4$ \citep{2015A&A...575A..74S}
should quench immediately to explain the number of the fast-quenching galaxies at $z\sim2$.
Although this is a rough estimate, 
to explain the number density of fast-quenching galaxies at $z=2$, 
it is not enough to account for so-called starbursting galaxies 
that have SFRs above three times the SFR$_{\rm MS}$. 
The simulations (e.g., \citealt{2004MNRAS.352..571B}) and those combined observations \citep{2020MNRAS.499.4748M}
also predicted that massive galaxies in the main sequence are the major progenitors of massive quiescent galaxies today.

\subsection{Morphologies of fast quenching galaxies}

Massive quiescent galaxies at up to $z=4.658$ have been discovered 
using the 10-m class ground-based telescopes and $JWST$ \citep{2023MNRAS.520.3974C,2023Natur.619..716C,2023ApJ...947...20V}
and raise a question about their formation and quenching scenario. 
Although there is a selection bias, most of them likely formed stars in a short timescale (a few 100 Myr or less)
according to the spectral analysis.
The fast-quenching galaxies selected by our methods are likely their low-z analogs.

The quiescent galaxies with a wide variety of quenching timescales and mechanisms 
have been reproduced by the EAGLE \citep{2015MNRAS.450.1937C,2015MNRAS.446..521S} and IllustrisTNG simulation \citep{2018MNRAS.475..624N}.
Like the observed ones, quiescent galaxies at high redshift tend to quench rapidly in these simulations.
AGN feedback is thought to be the plausible quenching mechanism for rapid quenching galaxies.
Indeed, several studies find quiescent galaxies with strong ionized gas outflows at $z>3$ 
\citep{2022ApJ...935...89K,2023Natur.619..716C}.
On the other hand, these latest simulations predict that 
the quenching timescales are not tightly related to the morphological transformation
\citep{2019MNRAS.485.2656C,2022MNRAS.515..213P}; 
about half the quiescent elliptical galaxies today are predicted 
to have disc morphologies at the epoch of quenching.
They are predicted to evolve into ellipticals through internal (secular evolution) or external (e.g., ram-pressure stripping, harassment, and minor mergers) processes after quenching. 
However, all of the recently quenched fast-quenching galaxies found in this study in Table \ref{tab:specz}, 
which should preserve the properties at the quenching, 
show elliptical-like compact morphologies.
Many of the $z>3$ quiescent galaxies found with $JWST$ 
are similarly characterized as compact ellipticals \citep{2023Natur.619..716C,2023arXiv230706994I}.
This suggests a link between quenching timescales and morphological transformation different from the above simulations or an interplay of their independent tendencies.

\section{Conclusions}

We constructed a new photometric selection method 
for fast $(\tau\leq0.1$ Gyr) quenching galaxies 
by detecting the spectral break at $\sim1600$ \AA~appearing when early A-type stars disappear. 
The robustness of this method for fast-quenching galaxies at $\sim2$ was successfully confirmed by 
the test with the photometric and spectroscopic catalog of galaxies in the COSMOS field. 
The main progenitors of such fast-quenching galaxies at $z=2$
are likely massive SFGs not far above the star-formation main sequence at $z=3-4$.
The fast-quenching galaxies selected here are plausible analogs 
of the most distant quiescent galaxies found with $JWST$.
By applying this selection method to the existing and future wide imaging surveys 
like the HSC-SSP, LSST, $WISE$, Euclid, and Roman surveys, 
we can survey the rare, most massive fast-quenching galaxies at $z\sim2$.
Lower mass ones can be selected in the several $JWST$ MIRI survey fields.
They can help understand the quenching mechanisms in the early Universe
by the statistical analysis and detailed spectral analysis of kinematics and stellar metallicities hardly studied at higher redshift even with $JWST$. 

\section*{Acknowledgements}

MK is supported by JSPS KAKENHI Grant Numbers  20K14530 \& 21H044902 and 
Tohoku University Center for Gender Equality Promotion (TUMUG).
The Hyper Suprime-Cam (HSC) collaboration includes the astronomical communities of Japan and Taiwan, and Princeton University. The HSC instrumentation and software were developed by the National Astronomical Observatory of Japan (NAOJ), the Kavli Institute for the Physics and Mathematics of the Universe (Kavli IPMU), the University of Tokyo, the High Energy Accelerator Research Organization (KEK), the Academia Sinica Institute for Astronomy and Astrophysics in Taiwan (ASIAA), and Princeton University. Funding was contributed by the FIRST program from the Japanese Cabinet Office, the Ministry of Education, Culture, Sports, Science and Technology (MEXT), the Japan Society for the Promotion of Science (JSPS), Japan Science and Technology Agency (JST), the Toray Science Foundation, NAOJ, Kavli IPMU, KEK, ASIAA, and Princeton University. 
This paper makes use of software developed for Vera C. Rubin Observatory. We thank the Rubin Observatory for making their code available as free software at http://pipelines.lsst.io/.
This paper is based on data collected at the Subaru Telescope and retrieved from the HSC data archive system, which is operated by the Subaru Telescope and Astronomy Data Center (ADC) at NAOJ. Data analysis was in part carried out with the cooperation of Center for Computational Astrophysics (CfCA), NAOJ. We are honored and grateful for the opportunity of observing the Universe from Maunakea, which has the cultural, historical and natural significance in Hawaii. 
This work is based on observations taken by the 3D-HST Treasury Program (GO 12177 and 12328) with the NASA/ESA HST, which is operated by the Association of Universities for Research in Astronomy, Inc., under NASA contract NAS5-26555.
This work is based on observations taken by the CANDELS Multi-Cycle Treasury Program with the NASA/ESA HST, which is operated by the Association of Universities for Research in Astronomy, Inc., under NASA contract NAS5-26555.
This paper makes use of the following ALMA data: ADS/JAO.ALMA\#2015.1.00379.S, ADS/JAO.ALMA\#2013.1.01292.S.
ALMA is a partnership of ESO (representing its member states), NSF (USA) and NINS (Japan), together with NRC (Canada) and NSC and ASIAA (Taiwan) and KASI (Republic of Korea), in cooperation with the Republic of Chile. The Joint ALMA Observatory is operated by ESO, AUI/NRAO, and NAOJ.
\section*{Data Availability}
The observational data used in this paper are all publicly available. 
We used the COSMOS2020 catalog \citep{2022ApJS..258...11W}, spectroscopic redshifts 
from \citet{2017ApJ...834...18B}, \citet{2020ApJ...888....4S} \& \citet{2022PASJ...74..247A}, spectra from 3D-HST\footnote{https://archive.stsci.edu/prepds/3d-hst/}, images from CANDELS survey\footnote{http://arcoiris.ucolick.org/candels/}, galaxy morphology catalog from \citet{2014ApJ...788...28V}, 
$Spitzer$ MIPS 24 $\mu$m image and catalog downloaded from the COSMOS survey\footnote{https://cosmos.astro.caltech.edu}, 
and ALMA data available via the ALMA archive\footnote{https://almascience.nrao.edu/aq/}.
The stellar population synthesis modeling was performed using {\sf CIGALE} \citep{2019A&A...622A.103B}\footnote{https://cigale.lam.fr}.
The overall data analysis was performed with {\sf Python} 3\footnote{https://www.python.org}.




\bibliographystyle{mnras}





\appendix

\section{Different combinations of filters}
\label{sec:diff_filter}
\begin{figure*}
\centering
\includegraphics[width=180mm]{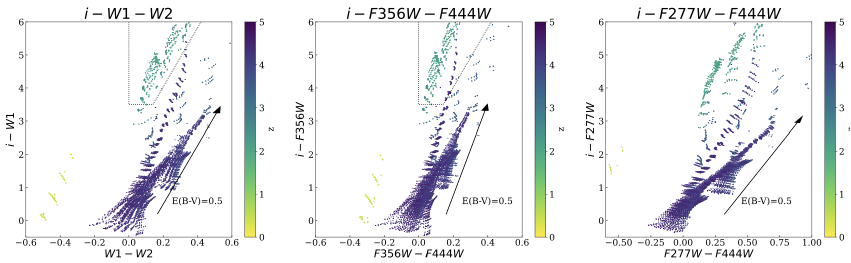}
    \caption{Similar to the $i-$ [3.6] vs. [3.6]$ - $[4.5] color diagram in Figure \ref{fig:color_model} but for other filter combinations. $i$-band is that of HSC on the Subaru Telescope. The panels are $i-W1$ vs. $W1-W2$ of $WISE$, and $i-F356W$ vs. $F356W-F444W$ and $i-F277W$ vs. $F277W-F444W$ of NIRCam on $JWST$ from left to right. The color criterion for $i$[3.6][4.5] selection is shown in the left and right color diagrams, which are the close analog of $i-$ [3.6] vs. [3.6]$ - $[4.5] color diagram.}
    \label{fig:others}
\end{figure*}

Figure \ref{fig:others} is similar to the $i-$ [3.6] vs. [3.6]$ - $[4.5] color diagram 
in Figure \ref{fig:color_model} but for other filter combinations. 
The panels are $i-W1$ vs. $W1-W2$ of $WISE$, and $i-F356W$ vs. $F356W-F444W$ and $i-F277W$ vs. $F277W-F444W$ using filters of NIRCam on $JWST$ for left to right.
The color sequences on the $i-W1$ vs. $W1-W2$ and $i-F356W$ vs. $F356W-F444W$ color diagrams
are almost the same as that on $i-$ [3.6] vs. [3.6]$ - $[4.5] color diagram.
The behavior on $i-F277W$ vs. $F277W-F444W$ is also almost similar to $i-$ [3.6] vs. [3.6]$ - $[4.5] color diagram.
 
\section{Different metallicities}
\label{sec:diff_metal}

\begin{figure*}
\centering
	\includegraphics[width=170mm]{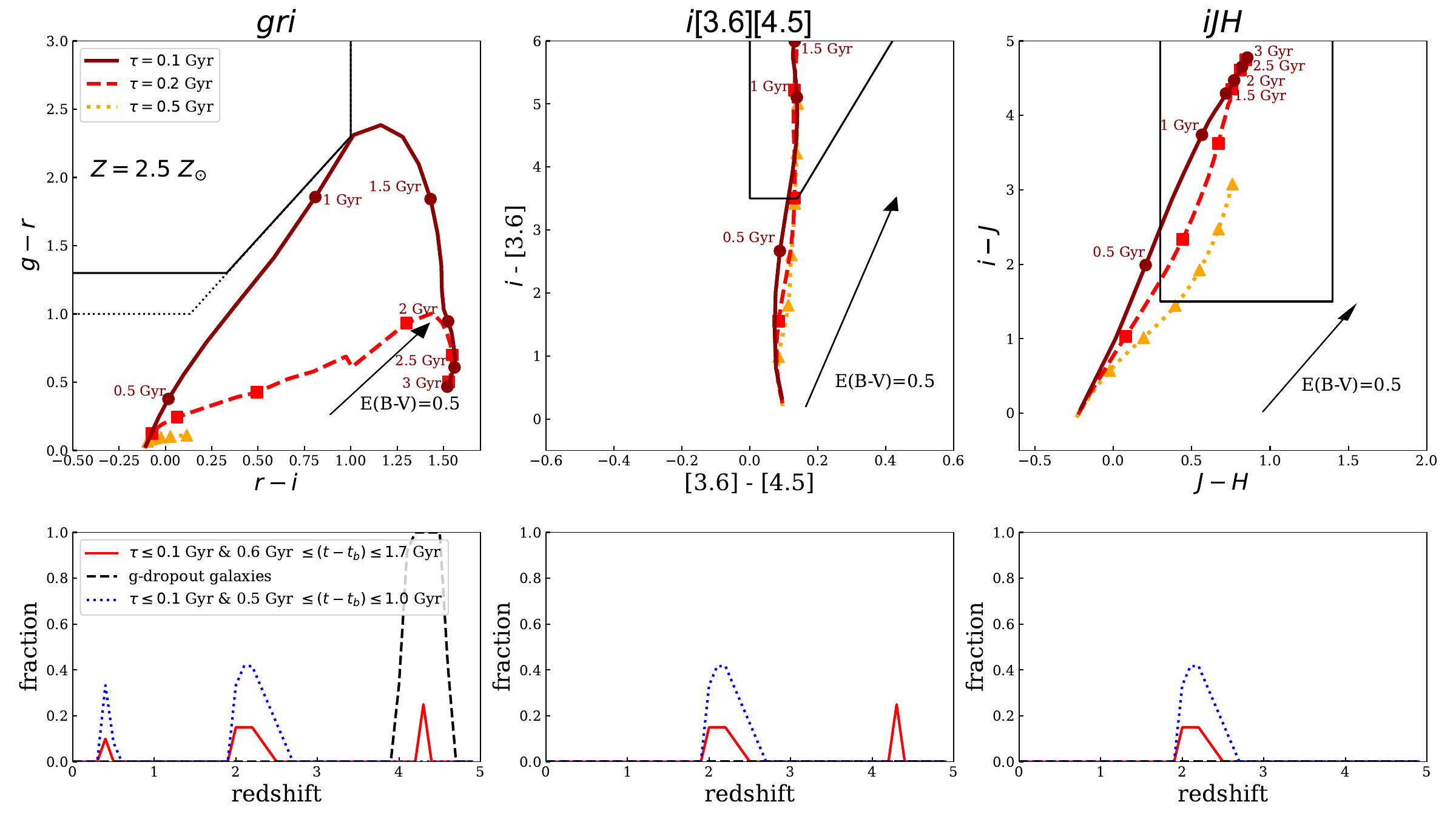}
    \caption{Similar to the central and right panels of Figure \ref{fig:color_model} but for the models with $Z = 2.5~Z_{\odot}$. In the bottom panels, in addition to the redshift selection functions in Figure \ref{fig:color_model}, the blue dotted curve shows that for models with $\tau\leq0.1$ Gyr and 0.5 $\leq (t -t_{\rm burst}$)/Gyr $\leq$ 1.0.}
 \label{fig:metal2p5Zsun}
\end{figure*}

\begin{figure*}
\centering
	\includegraphics[width=170mm]{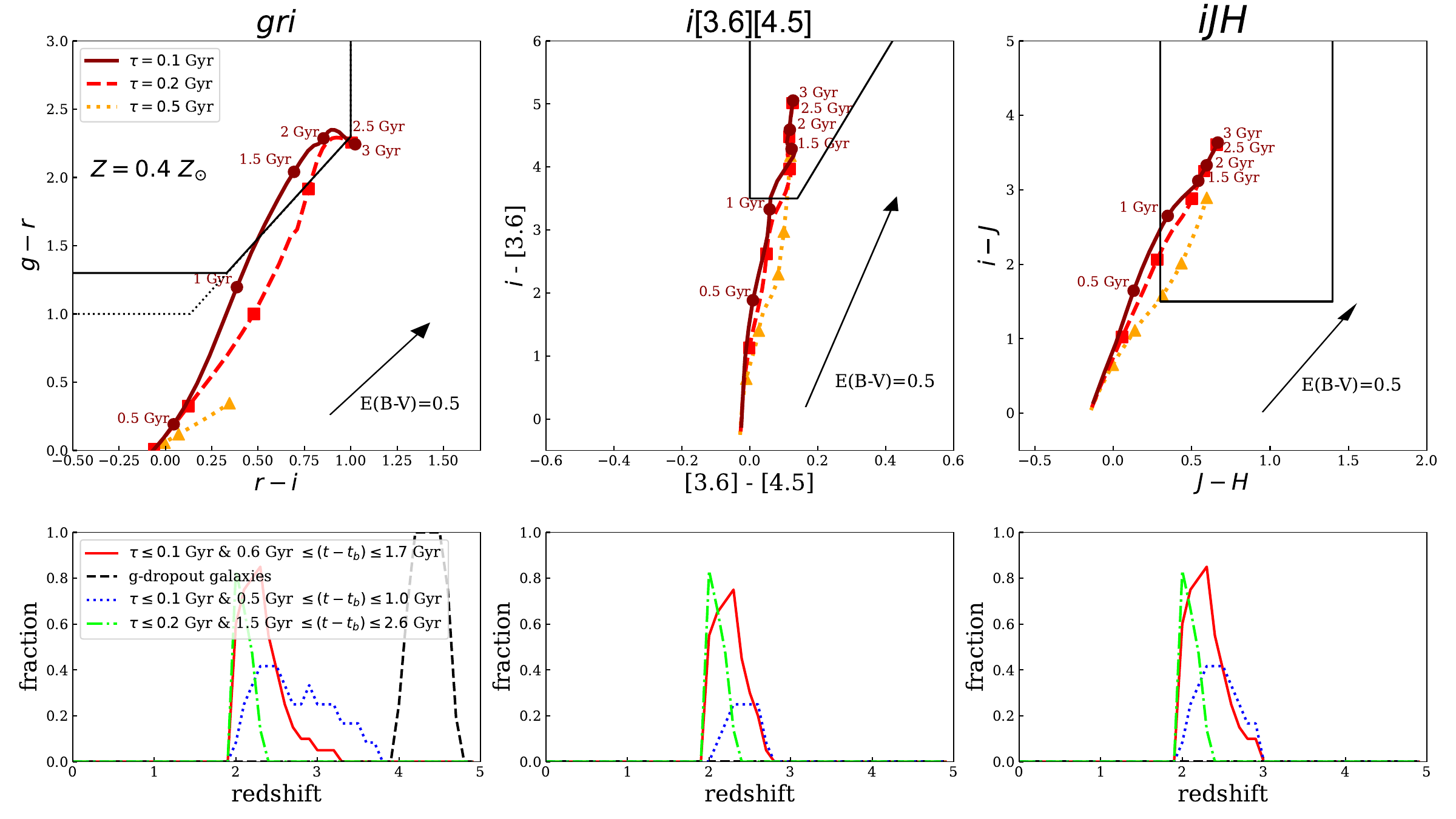}
	    \caption{Similar to Figure \ref{fig:metal2p5Zsun} but for the models with $Z = 0.4~Z_{\odot}$. The lime dashed-dotted curve in the bottom panels shows the redshift selection function for models with $\tau\leq0.2$ Gyr and 1.5 $\leq (t -t_{\rm burst}$)/Gyr $\leq$ 2.6.}
    \label{fig:metal0p4Zsun}
    \end{figure*}

\begin{figure*}
\centering
	\includegraphics[width=170mm]{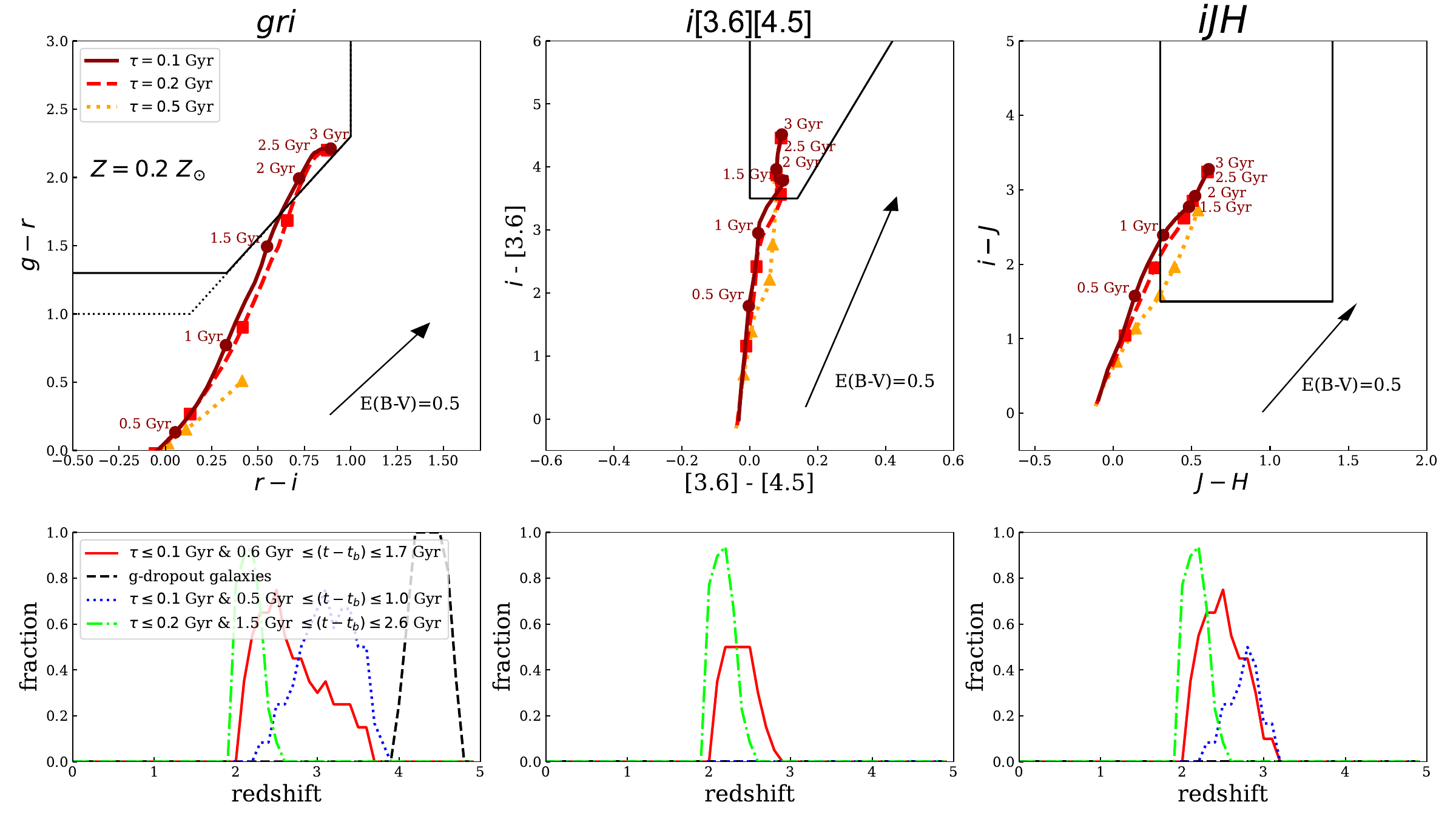}
    \caption{Similar to Figure \ref{fig:metal0p4Zsun} but for the models with $Z = 0.2~Z_{\odot}$. The lime dashed-dotted curve in the bottom panels shows the redshift selection function for models with $\tau\leq0.2$ Gyr and 1.5 $\leq (t -t_{\rm burst}$)/Gyr $\leq$ 3.0.}
    \label{fig:metal0p2Zsun}
\end{figure*}

In this section, we describe models with different metallicities.
Figure \ref{fig:metal2p5Zsun} to \ref{fig:metal0p2Zsun} are similar to Figure \ref{fig:color_model}, 
but for $Z = 2.5, 0.4~Z_{\odot}$ and $0.2~Z_{\odot}$.
The top and bottom panels show the model color tracks and redshift selection functions 
like the central and right panels of Figure \ref{fig:color_model}, respectively.
The ranges of the models satisfying the color criteria differ with metallicities.
Fast-quenching galaxies with with $\tau\leq0.1$ Gyr and 0.6 $\leq (t-t_{\rm burst})$/Gyr $\leq$ 1.7 
are similarly selected well for $Z=0.2~Z_{\odot}$ and 0.4 $Z_{\odot}$ while they are merely selected for $Z = 2.5 ~Z_{\odot}$.
The blue-dotted curves in Figure \ref{fig:metal2p5Zsun} to \ref{fig:metal0p2Zsun} bottom panels
show the redshift selection functions for the models 
with $\tau\leq0.1$ Gyr and 0.5 $\leq (t -t_{\rm burst}$)/Gyr $\leq$ 1.0, and $\tau\leq0.2$ Gyr.
The green dashed-dotted curves in Figure \ref{fig:metal0p4Zsun} and \ref{fig:metal0p2Zsun} bottom panels
show the redshift selection functions for the models  
with $\tau\leq0.2$ Gyr, and 1.5 $\leq  (t -t_{\rm burst}$)/Gyr$\leq$ 2.6 and 1.5 $\leq (t -t_{\rm burst}$)/Gyr $\leq$ 3.0, respectively.
In the case of $Z=2.5~Z_{\odot}$, less $(t-t_{\rm burst})$ models, like so-called post-starburst galaxies, at $z\sim2$ are preferentially selected by the $gri$ color criterion.
In the case of $Z=0.2~Z_{\odot}$ and 0.4 $Z_{\odot}$, less $(t-t_{\rm burst}) $ models at $z\sim3$ 
and longer $(t-t_{\rm burst})$ models at $z\sim2$ are also selected.
Note that longer $(t-t_{\rm burst})$ objects should have less chance to be detected because the UV emission of galaxies diminishes with time after quenching, and fewer galaxies have been quenched at higher redshift.
To summarize, our color criteria likely work well for fast-quenching galaxies with solar to sub-solar metallicities. 

\bsp	
\label{lastpage}
\end{document}